# Phase transition and anomalous scaling in the quantum Hall transport of topological insulator Sn-Bi$_{1.1}$Sb$_{0.9}$Te$_2$S devices

Faji Xie[1], Shuai Zhang[1], Qianqian Liu[1], Chuanying Xi[2], Ting-Ting Kang[4], Rui Wang[1], Boyuan Wei[1], Xing-Chen Pan[1], Minhao Zhang[1], Fucong Fei[1], Xuefeng Wang[3], Li Pi[2], Geliang L. Yu[1], Baigeng Wang[1,†] and Fengqi Song[1,†]

[1]*National Laboratory of Solid State Microstructures, College of Physics and Collaborative Innovation Center of Advanced Microstructures, Nanjing University, Nanjing 210093, China*

[2]*High Magnetic Field Laboratory, Chinese Academy of Sciences and Collaborative Innovation Center of Advanced Microstructures, Hefei 230031, China*

[3]*National Laboratory of Solid State Microstructures, School of Electronic Science and Engineering and Collaborative Innovation Center of Advanced Microstructures, Nanjing University, Nanjing 210093, China*

[4]*National Laboratory for Infrared Physics, Shanghai Institute of Technical Physics, Chinese Academy of Sciences, Shanghai 200083, China*

† Corresponding authors. Emails: songfengqi@nju.edu.cn, bgwang@nju.edu.cn. Fax: +86-25-83595535. The first two authors contributed equally.

**Abstract**

The scaling physics of quantum Hall transport in optimized topological insulators with a plateau precision of ~1/1000 $e^2/h$ is considered. Two exponential scaling regimes are observed in temperature-dependent transport dissipation, one of which accords with thermal activation behavior with a gap of 2.8 meV (> 20 K), the other being attributed to variable range hopping (1-20 K). Magnetic field-driven plateau-to-plateau transition gives scaling relations of $(dR_{xy}/dB)^{max} \propto T^{-\kappa}$ and $\Delta B^{-1} \propto T^{-\kappa}$ with a consistent exponent of $\kappa \sim 0.2$, which is half the universal value for a conventional two-dimensional electron gas. This is evidence of percolation assisted by quantum tunneling, and reveals the dominance of electron-electron interaction of the topological surface states.

The quantum Hall (QH) effect is seen in two-dimensional (2D) electron systems perpendicular to a magnetic field, in which the continuous density of states is broken, leading to discrete Landau levels (LLs), and reflecting the nature of topological quantum numbers. QH systems are ideal for studying transport phenomena thanks to their clear physical characteristics, allowing the study of a number of important issues, such as quantum metrology and scaling law [1]. Among these issues, QH plateau-to-plateau transition (PPT) behavior is of great importance in terms of an interesting transition from the localization to the delocalization state in 2D systems [2, 3].

The transition between adjacent quantized QH plateaus occurs through changes to other parameters, such as the magnetic field ($B$) or the Fermi energy ($E_F$) [2, 3]. In the PPT regime, the degree of delocalization is characterized by a localization length ($\xi$) that decays exponentially from the quantum critical points. At infinite temperatures, temperature-dependent PPT behavior results in a power-law relationship according to finite scaling theory. The half width ($\Delta B$) of the peaks shrinks as the temperature ($T$) decreases according to the scaling function $\Delta B^{-1} \propto T^{-\kappa}$, and the maximum of $dR_{xy}/dB$ diverges as $(dR_{xy}/dB)^{\max} \propto T^{-\kappa}$. The exponent $\kappa$ can be expressed as $\kappa = p/2\gamma$, where $p$ is the temperature exponent of the inelastic scattering length or quantum coherence length, and $\gamma$ is the localization length exponent. The universal critical exponent $\kappa = 0.42$ is established for a conventional 2D electron gas (2DEG) [2-8]. This can be understood in terms of the quantum percolation, involving the quantum tuning between chiral edge modes around a single droplet [9].

Ever since the discovery of topological insulators (TIs) [10-14], topological surface states (TSSs), being a kind of 2D helical electron system with linear dispersion and protection from backscattering, have attracted a great deal of interest. Various intriguing transport phenomena of TSSs have been observed experimentally, including weak antilocalization (WAL) [15, 16], spin–orbit torque, [17] and others. The QH effect [18-22] of TSSs is very important, being a half-integer QH effect, different from conventional 2DEG. In this 2D TSS system, there is some significance attached to the investigation of scaling behavior with a topological electron.

In this letter, we describe an optimized TI [23, 24] Sn-Bi$_{1.1}$Sb$_{0.9}$Te$_2$S (Sn-BSTS), which gives a mobility in some devices of up to ~10000 cm$^2$/Vs, and which is an ideal 2D electron system with pure TSS, allowing us to study topological quantum transport. Because such a high-quality QH effect is observed, the dominant mode of transport dissipation in the QH state changes from thermal activation to variable range hopping (VRH) at a critical temperature $T_c$ ~ 20 K. At temperatures below 20 K, the PPT behavior produces an exponent $\kappa$ ~ 0.2, nearly half the universal value in conventional 2DEG. To allow further discussion on the scattering mechanism of TSS transport, we investigate the dephasing behavior and obtain a temperature coefficient of inelastic scatting $p$ ~ 1.1, which implies that the electron interactions follow a predominantly inelastic scattering mechanism, consistent with our critical exponent.

To study the scaling physics in a 2D electron system of TSSs, high quality TIs are required. We grew high quality Sn-BSTS crystals by melting high-purity elements of bismuth, antimony, tellurium and sulfur in a molar ratio of 1.1:0.9:2:1, and a small quantity of tin powder. An optical photograph of the crystals is shown in Fig. 1(a). Angle-resolved photoemission spectroscopy (ARPES) of the bulk crystal was performed at 10 K, as shown in Fig. 1(b). A Dirac cone with linear dispersion can clearly be seen at the center of the Brillouin zone. The Dirac point (DP) is about 126 meV below the Fermi level, indicating *n*-type carriers.

Using mechanical exfoliation, Sn-BSTS nanoflakes were transferred to $SiO_2$/Si substrates, before being fabricated into gated Hall-bar devices. A typical atomic force microscopy graph of the device is shown in the inset of Fig. 1(e). At high temperatures, the resistances of all the devices with different thicknesses (*t*) exhibit insulating behavior, as shown in Fig. 1(c). When $T < 125$ K, they show metallic behavior, implying that TSS is dominant. Hence, pure TSSs exist at a large scale at finite temperatures, which allows us to investigate some intriguing transport mechanisms in 2D topological electron systems.

We first observed the low-dissipation QH states of the TSS in Sn-BSTS devices. Figure 1(d) shows the QH effect measured at $T = 6$ K in Sample A ($t = 55$ nm). The highest Hall mobility ($\mu$) is 10277 $cm^2$/Vs when the Fermi surface is tuned close to the DP of the bottom surface, with a corresponding carrier density $n_e = 2 \times 10^{11}$ $cm^{-2}$. The inset of Fig. 1(d) shows that the accuracy of the Hall plateau resistance with $h/e^2$ (*h* is Planck constant, and *e* is the elementary charge) is more than 99.4％ at $B = 4$ T. The high-quality QH plateau originates from the quantized LLs and offers convenience in the measurement of the QH edge states with nonlocal contacts. The phase transition between integer $v = 1$ and 2 QH states in Sample B ($t = 62$ nm) is shown in Figure 1(e). The $v = 1$ QH state shows a giant plateau up to 38 T, which is also observed in graphene [25]. From the nonlocal measurement shown in Fig. 1(f), we further confirm the existence of dissipationless edge conduction for the QH states. The inset shows the contact serial number. Contacts 1 and 2 are designated as the source or drain pads in two-terminal measurement ($R_{12/12}$), while contacts 5 and 4 are used as additional voltage probes ($R_{54/12}$). In the QH region, the dissipationless current flow follows the contacts in the order 1→6→5→4→3→2 and "shorts" the path. As a result, $R_{54/12}$ drops close to zero when $R_{12/12}$ gives $h/ve^2$ with $v = 1$ or 2. This indicates the well-developed QH edge channels of the TSSs.

In QH systems, both the deviation of the Hall plateau and the residual longitudinal conductivity indicate dissipation of the quantum transport. The mechanism of transport dissipation in the QH states of the TSS has yet to be studied in detail, nevertheless we discuss the temperature-dependent QH states here with some care. The gate-tuned QH states in the highly quantized state are measured in Sample B with temperature increasing over a large range from 2 K to 40 K, as shown in Figs. 2(a, b). QH states for $v = -2, -1$ and 0 show plateaus, interpreted as $v_{\text{bottom}}$ varying from -3/2, -1/2 to +1/2 by tuning the Fermi level while $v_{\text{top}} = -1/2$. As noted in Figs. 2(a, b), the QH states can survive at temperatures of more than 40 K, the $v = -1$ plateau deviation is still close to $-e^2/h$, and the dip of $\sigma_{xx}$ is close to 0. Figure 2(e) shows the renormalization group flow diagram of the QH states varying with changing Fermi surface and different temperatures in ($\sigma_{xy}$, $\sigma_{xx}$) space [26, 27]. The flow in ($\sigma_{xy}$, $\sigma_{xx}$) tends to converge towards ($-e^2/h$, 0) and (0, 0), which corresponds to the $v = -1$ and 0 QH states, respectively.

In the quantized regime, we find that dissipation of the edge states follows two behaviors at different temperature regimes. The temperature dependence of the $v = -1$ QH state at higher temperatures reveals the minimum value of $\sigma_{xx}$, pointing to the thermal activation of the transport dissipation [1], varying as $\exp(-\Delta E_0 / k_B T)$ with an activation gap $E_0 \sim 2.8$ meV, as shown in Fig. 2(c). We also find that $\sigma_{xx}(T)$ and $\Delta\sigma_{xy}(T)$ [$= \sigma_{xy}(T) - \sigma_{xy}(0)$] exhibit a linear relationship (red dashed line) in Fig. 2(f), which indicates that both undergo the same ratio of the energy gaps in the dominant region of thermal activation, similar to that in BiSbTeSe$_2$ [18]. When the temperature is below 20 K, the minimum $\sigma_{xx}$ drops more slowly and begins to deviate from the thermal activation fitting curve, suggesting that the mechanism has been frozen, and the VRH dominates the dissipation of the QH edge states. While in the VRH region, the relationship is given by

$$\sigma_{xx} \propto \sigma_0 \exp(-\sqrt{T_0 / T}), \quad (1)$$

where $\sigma_0 \propto 1/T$ and $T_0 \propto e^2 / \varepsilon \xi$ ($\varepsilon$ is the dielectric constant). In Fig. 2(d), the linear fitting of Eq. (1) matches the experimental data very well when $T < 20$ K, meaning that VRH is dominant. In Fig. 2(f), the blue dashed line indicates that $\sigma_{xx}$ and $\Delta\sigma_{xy}$ share the same localization radius in the VRH region. The two fitting lines with different slopes (the red and blue dashed lines) intersect at a single point for a temperature of 20 K in Fig. 2(f), suggesting a crossover from thermal activation to VRH. Therefore, there is a transition from thermal activation to VRH at a critical temperature of 20 K. This indicates that the transport of TSSs enters a low-dissipation regime of hopping conductivity, which allows us to study the scaling mechanism of a 2D topological electron system over a large temperature range.

The QH effect is a transport of edge states in a strong Anderson localization regime. The QH plateaus represent separate energy regions of localized states, while between two adjacent plateaus there is a specific critical field $B_c$ associated with one energy level of the extended states. The PPT is therefore a localization-delocalization transition. From the idea of quantum phase transitions, there is a power law divergence of the localization length $\xi \propto |\delta|^{-\gamma}$, where $\delta = B - B_c$ (or $E - E_c$). It is generally believed that the localization length exponent $\gamma$ is universal, and $\gamma \sim 2.4$ [7-9, 28-32]. At finite temperatures, this $\xi \propto |\delta|^{-\gamma}$ power law can be translated using finite-size scaling theory into a temperature-scaling form, $(dR_{xy}/dB)^{max} \propto T^{-\kappa}$ and $\Delta B^{-1} \propto T^{-\kappa}$. Our precise study of PPT behavior in a 2D topological electron system, as shown in Figures 3(a, b), reveals a transition from the plateau of filling factors $v = 2$ to $v = 1$ over a temperature range from 5 K to 16 K. The values of the maxima of $dR_{xy}/dB$ and the half width $\Delta B$ of the $R_{xx}$ peak are at different temperatures, as extracted from the experimental data shown in Fig. 3(d). The slope of the dashed straight lines gives $(dR_{xy}/dB)^{max} \propto T^{-\kappa}$ and $\Delta B^{-1} \propto T^{-\kappa}$ with $\kappa = 0.21\pm0.02$ and $0.20\pm0.01$, respectively. We also carried out some nonlocal measurements to observe localization-delocalization transition behavior by tuning the Fermi energy. As shown in Fig. 1(f), the peak between adjacent dips of $R_{54/12}$ also displays the PPT of the $v = 1$ and 2 QH states. Following the power law of $\xi \propto |E - E_c|^{-\gamma}$, it is useful to derive a temperature scaling form with $\Delta V_g^{-1} \propto T^{-\kappa}$. Figure 3(c) shows $R_{54/12}$ at $B = 12$ T by changing $T$ from 2 K to 16 K. The half width of the nonlocal resistance peaks gives $\Delta V_\text{g}^{-1} \propto T^{-\kappa}$ with $\kappa = 0.20\pm0.015$, as shown in Fig. 3(d). This is the first observation of the universal scaling law of PPT in TSSs. We note a similar scaling law in the quantum anomalous Hall (QAH) effect experiment [33], which may indicate the nature of the QAH state.

In conventional 2DEG, the critical exponent has been found to be universal and $\kappa = 0.42$ [6-8], with $\gamma \sim 2.4$ and $p = 2$. However, other experiments show that the exponent is not universal and deviates from 0.42. This discrepancy could be a result of various factors, including long-range disorder [7], spin-degenerate LLs [34], and macroscopic disorder [2], among others. For example, $\kappa$ will increase from 0.42 towards 0.75 in conventional 2DEG when the disorder changes from short to long range [7], inducing a crossover from quantum to classical percolation with a change in $\gamma$ from 7/3 to 4/3 [35]. None of these reasons explains our result, however, which is nearly half the universal value in conventional 2DEG. The long-range disorder only leads to a value greater than 0.42, and the TSSs are not spin-degenerate. The discrepancy is therefore an intrinsic distinction between TSSs and conventional 2DEG.

We find that a temperature-dependent scaling law with a different exponent $\kappa$ can result from the influence of the dynamically altered localization length with a different inelastic scattering coefficient $p$ [36-38]. The WAL effect, shown in Fig. 4(a), provides a reliable means of studying electron dephasing in 2D TSSs. The magnetoconductivity at low field can be described by the Hikami-Larkin-Nagaoka

(HLN) equation,

$$\Delta\sigma(B) = -\alpha \frac{e^2}{\pi h} \left[ \psi\left(\frac{1}{2} + \frac{B_\phi}{B}\right) - \ln\left(\frac{B_\phi}{B}\right) \right], \qquad (2)$$

where $\psi(x)$ is a digamma function, and $B_\phi = \hbar/4el_\phi^2$ is the dephasing field. By fitting with the HLN formula, the dephasing length $l_\phi$ of the TSSs can be extracted, and is much longer than the device thickness. According to the current framework, electron-electron interaction and electron-phonon interaction are the two main sources of the electron phase relaxation. While electron-phonon interaction usually gives $p = 2\sim 4$ [39], electron-electron interaction leads to $p = 2/3$ and 1 for 1D and 2D, respectively [40, 41]. Electron-electron interaction with small energy tranfers is called Nyquist dephasing. Previous studies of APERS [42] and transport [43] indicate that the electron-phonon interaction occurs on the TSSs, because a vast bulk phonon "sea" affects the TSS electrons. However, this is not the case for our samples.

Figure 4(b) shows that the temperature-dependent $l_\phi$ has a power-law dependence $l_\phi \propto T^{-p/2}$ with $p \sim 1.1$ at low temperatures, while for the QH experiments in conventional 2DEG, the result is $p = 2$ [6]. This means that at low temperatures, the transport in TI almost entirely originates from the TSSs, and the $l_\phi$ behavior confirms the Nyquist dephasing. Therefore, the value of $p$ indicates that the dominant inelastic scattering mechanism is electron-electron interactions in the TSSs, which in turn suggests that no itinerant electrons exist in the bulk for the optimized TIs, and no electron-phonon interaction survives, which is the same as the result for graphene [44] on boron nitride substrate with no disturbances from phonons. Therefore, the TSSs in high-quality TIs can be protected effectively from contamination from bulk phonons.

Furthermore, from the foregoing analysis the exponent $\kappa$ in the TSSs should indeed be nearly half the value in conventional 2DEG, consistent with the result seen in our experiments.

In conclusion, we observed well-developed QH states of TSSs in high-quality TI Sn-BSTS devices. When the temperature is below 20 K, the transport is dominated by VRH, which is a low dissipation QH regime. By considering the temperature-dependent QH transition in this 2D topological electron system, we find a universal scaling relationship with $\kappa \sim 0.20$ in both the local and nonlocal measurements. This is due to the fact that the electron-electron interaction dominates the TSSs transport, indicated by the dephasing behavior. Our investigation reveals novel topological scaling physics, and opens the way for new perspectives on transport in TIs.

This work was supported by National Key R&D Program of China (Grant No. 2017YFA0303200), the National Natural Science Foundation of China (Grant Nos. U1732273, U1732159, 91622115, 11522432 and 11574217), the Natural Science Foundation of Jiangsu Province (Grant No. BK20160659). We also acknowledge the assistance of the Nanofabrication and Characterization Center at the Physics College of Nanjing University. We would like to thank Professor Tigran Sedrakyan from University of Massachusetts Amherst, Professor Haizhou Lu from the Southern University of Science and Technology of China, and Professor Jing Wang from Fudan University, for some very stimulating discussions.

# References

[1] M. E. Cage, K. Klitzing, A. Chang, F. Duncan, M. Haldane, R. B. Laughlin, A. Pruisken and D. Thouless, *The quantum Hall effect*. (Springer Science & Business Media, 2012).

[2] B. Huckestein. Rev. Mod. Phys. **67**, 357-396 (1995).

[3] S. L. Sondhi, S. M. Girvin, J. P. Carini and D. Shahar. Rev. Mod. Phys. **69**, 315-333 (1997).

[4] H. P. Wei, D. C. Tsui, M. A. Paalanen and A. M. Pruisken. Phys. Rev. Lett. **61**, 1294-1296 (1988).

[5] A. M. Pruisken. Phys. Rev. Lett. **61**, 1297-1300 (1988).

[6] H. P. Wei, L. W. Engel and D. C. Tsui. Phys. Rev. B **50**, 14609-14612 (1994).

[7] W. Li, G. A. Csathy, D. C. Tsui, L. N. Pfeiffer and K. W. West. Phys. Rev. Lett. **94**, 206807 (2005).

[8] W. Li, C. L. Vicente, J. S. Xia, W. Pan, D. C. Tsui, L. N. Pfeiffer and K. W. West. Phys. Rev. Lett. **102**, 216801 (2009).

[9] J. T. Chalker and P. D. Coddington. J. Phys. C: Solid State Phys. **21**, 2665-2679 (1988).

[10] M. Z. Hasan and C. L. Kane. Rev. Mod. Phys. **82**, 3045-3067 (2010).

[11] J. E. Moore. Nature **464**, 194-198 (2010).

[12] X.-L. Qi and S.-C. Zhang. Rev. Mod. Phys. **83**, 1057 (2011).

[13] Y. Ando. J. Phys. Soc. Jpn. **82**, 102001 (2013).

[14] H. Zhang, C.-X. Liu, X.-L. Qi, X. Dai, Z. Fang and S.-C. Zhang. Nat. Phys. **5**, 438 (2009).

[15] J. Chen, *et al.* Phys. Rev. Lett. **105**, 176602 (2010).

[16] H.-T. He, G. Wang, T. Zhang, I.-K. Sou, G. K. L. Wong, J.-N. Wang, H.-Z. Lu, S.-Q. Shen and F.-C. Zhang. Phys. Rev. Lett. **106**, 166805 (2011).

[17] A. R. Mellnik, *et al.* Nature **511**, 449 (2014).

[18] Y. Xu, I. Miotkowski, C. Liu, J. Tian, H. Nam, N. Alidoust, J. Hu, C.-K. Shih, M. Z. Hasan and Y. P. Chen. Nat. Phys. **10**, 956-963 (2014).

[19] R. Yoshimi, A. Tsukazaki, Y. Kozuka, J. Falson, K. S. Takahashi, J. G. Checkelsky, N. Nagaosa, M. Kawasaki and Y. Tokura. Nat. Commun. **6**, 6627 (2015).

[20] S. Zhang, *et al.* Nat. Commun. **8**, 977 (2017).

[21] C. Li, B. de Ronde, A. Nikitin, Y. Huang, M. S. Golden, A. de Visser and A. Brinkman. Phys. Rev. B **96**, 195427 (2017).

[22] N. Koirala, *et al.* Nano Lett. **15**, 8245-8249 (2015).

[23] B. Wu, X.-C. Pan, W. Wu, F. Fei, B. Chen, Q. Liu, H. Bu, L. Cao, F. Song and B. Wang. Appl. Phys. Lett. **113**, 011902 (2018).

[24] S. K. Kushwaha, *et al.* Nat. Commun. **7**, 11456 (2016).

[25] Z. R. Kudrynskyi, *et al.* Phys. Rev. Lett. **119**, 157701 (2017).

[26] J. G. Checkelsky, R. Yoshimi, A. Tsukazaki, K. S. Takahashi, Y. Kozuka, J. Falson, M. Kawasaki and Y. Tokura. Nat. Phys. **10**, 731-736 (2014).

[27] D. E. Khmelnitskii. Jetp Lett. **38**, 552-556 (1983).

[28] B. Huckestein and B. Kramer. Phys. Rev. Lett. **64**, 1437-1440 (1990).

[29] G. V. Milnikov and I. M. Sokolov. Jetp Lett. **48**, 536-540 (1988).

[30] D. H. Lee, Z. Q. Wang and S. Kivelson. Phys. Rev. Lett. **70**, 4130-4133 (1993).

[31] A. W. W. Ludwig, M. P. A. Fisher, R. Shankar and G. Grinstein. Phys. Rev. B **50**, 7526-7552 (1994).

[32] L. W. Engel, D. Shahar, C. Kurdak and D. C. Tsui. Phys. Rev. Lett. **71**, 2638-2641 (1993).

[33] X. Kou, *et al.* Nat. Commun. **6**, 8474 (2015).

[34] S. W. Hwang, H. P. Wei, L. W. Engel, D. C. Tsui and A. M. M. Pruisken. Phys. Rev. B **48**, 11416-11419 (1993).

[35] N. A. Dodoo-Amoo, K. Saeed, L. H. Li, E. H. Linfield, A. G. Davies and J. E. Cunningham. J. Phys.: Conf. Ser. **456**, 012007 (2013).

[36] J. Wang, B. Lian and S.-C. Zhang. Phys. Rev. B **89**, 085106 (2014).

[37] B. Huckestein and M. Backhaus. Phys. Rev. Lett. **82**, 5100-5103 (1999).

[38] D.-H. Lee, Z. Wang and S. Kivelson. Phys. Rev. Lett. **70**, 4130-4133 (1993).

[39] J. J. Lin and J. P. Bird. J. Phys.: Condens. Matter **14**, R501 (2002).

[40] B. L. Altshuler, A. G. Aronov and D. E. Khmelnitsky. J. Phys. C: Solid State Phys. **15**, 7367 (1982).

[41] J. Liao, Y. Ou, H. Liu, K. He, X. Ma, Q. K. Xue and Y. Li. Nat. Commun. **8**, 16071 (2017).

[42] X. Zhu, L. Santos, C. Howard, R. Sankar, F. C. Chou, C. Chamon and M. El-Batanouny. Phys. Rev. Lett. **108**, 185501 (2012).

[43] M. V. Costache, I. Neumann, J. F. Sierra, V. Marinova, M. M. Gospodinov, S. Roche and S. O. Valenzuela. Phys. Rev. Lett. **112**, 086601 (2014).

[44] F. V. Tikhonenko, A. A. Kozikov, A. K. Savchenko and R. V. Gorbachev. Phys. Rev. Lett. **103**, 226801 (2009).

**Figure Captions**

**Figure 1. Quantum Hall transport with low dissipation in optimized topological insulators.**

(a) Optical photograph of Sn-BSTS single crystals. The white scale bar is 5 mm. (b) The angle-resolved photoemission spectroscopy spectra of Sn-BSTS, which shows a Dirac cone. (c) The typical temperature-dependent sheet resistances in six devices with different thicknesses. (d) Magnetic field dependence of $R_{xx}$ and $R_{xy}$ of sample A measured at $V_g = 0$ V and $T = 6$ K. The inset shows an enlarged view of the plateau. They exhibit quantum Hall (QH) states with low dissipation. (e) The magnetic field dependent $R_{xx}$ (red line) and $R_{xy}$ (blue line) at 1 K and $V_g = 30$ V in sample B. The inset shows an atomic force microscopy picture of the device, whose thickness is 62 nm. (f) Back-gate voltage-dependent $R_{54/12}$ and $R_{12/12}$ in nonlocal measurements at 2 K and -12 T, which also exhibits QH edge states. The inset shows the measurement configuration.

**Figure 2. Transition from thermal activation to variable range hopping of the temperature dependent scaling of the QH dissipation (Sample B).**

(a) Back gate voltage-dependent $\sigma_{xy}$ at 12 T for different temperatures. (b) $\sigma_{xx}$ at 12 T for different temperatures. (c) Minima of $\sigma_{xx}$ (in log scale) as a function of $1/T$ at $\nu = 1$ plateau. The red dashed line indicates a linear fit at high temperature with thermal activation formula. (d) The variable range hopping (VRH) fitting by Eq. (2). The axes are rescaled to show a straight line. The critical temperature between thermal activation and VRH is about 20 K. (e) The renormalization group flow diagram in ($\sigma_{xy}$, $\sigma_{xx}$) space. (f) $\Delta\sigma_{xy}(T)$ versus minima of $\sigma_{xx}(T)$. Red and blue dashed lines indicate the linear fitting of different temperature regimes.

**Figure 3. The magnetic field-driven QH plateau-plateau transitions and anomalous scaling constant.**

(a) Magnetic field dependent $R_{xy}$ measured at zero gate voltage. (b) The $R_{xx}$ at zero gate voltage. (c) The back gate tuned $R_{54/12}$ at 12 T for different temperatures. (d) Blue: temperature dependence of the maxima of $dR_{xy}/dB$. The slope of the straight line gives $\left(dR_{xy}/dB\right)^{max} \propto T^{-\kappa}$ with $\kappa = 0.21\pm0.02$. Green: the half width of the longitudinal resistance peaks gives $\Delta B^{-1} \propto T^{-\kappa}$ with $\kappa = 0.20\pm0.01$. Purple: the half width of nonlocal resistance peaks gives $\Delta V_g^{-1} \propto T^{-\kappa}$ with $\kappa = 0.20\pm0.015$. Red dashed lines show linear fits.

**Figure 4. the temperature dependent weak antilocalization and its scaling physics.**

(a) The magnetoconductivity at different temperatures, which displays a weak antilocalization effect. (b) The coherence length $l_\phi$ of TSS extracted from the Hikami-Larkin-Nagaoka fitting at different temperatures, the red dashed line indicates the fitting of $l_\phi \propto T^{-p/2}$ with $p \approx 1.1$.

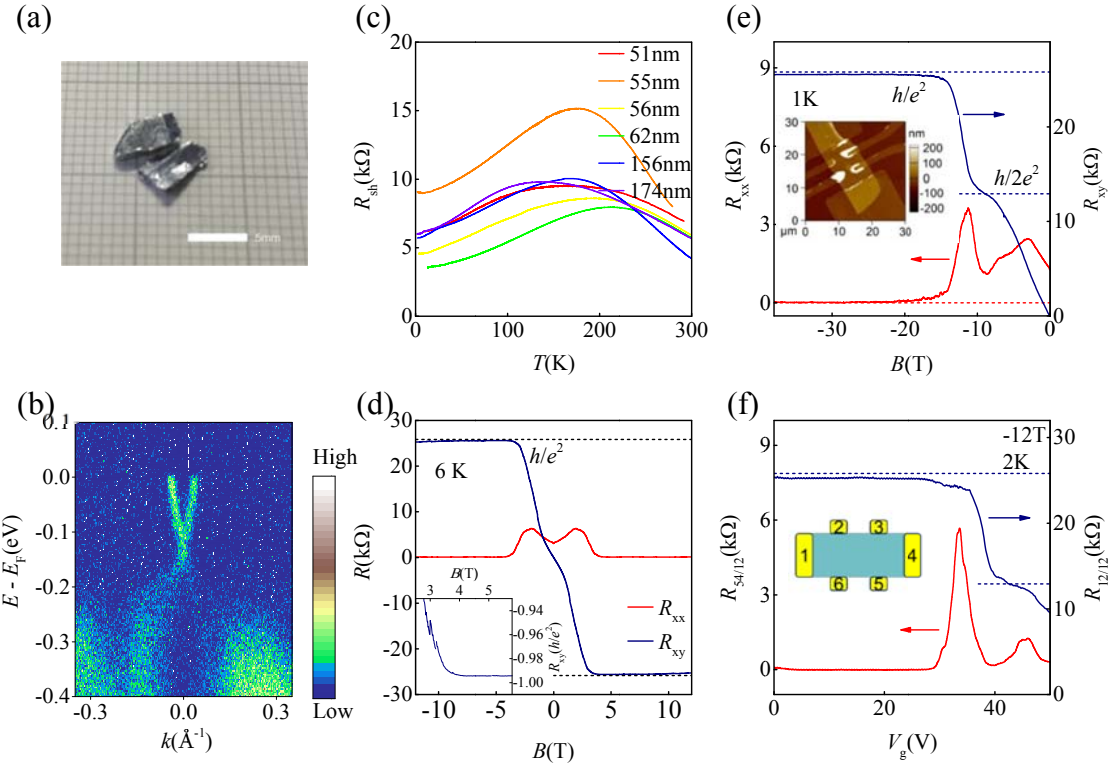

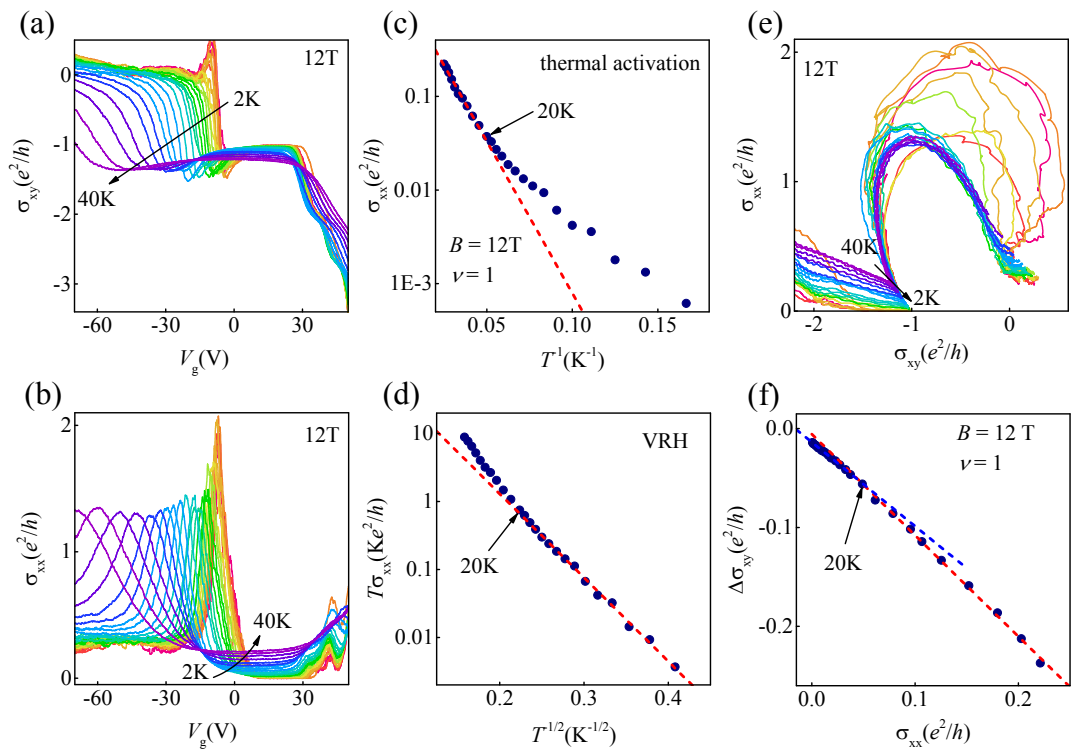

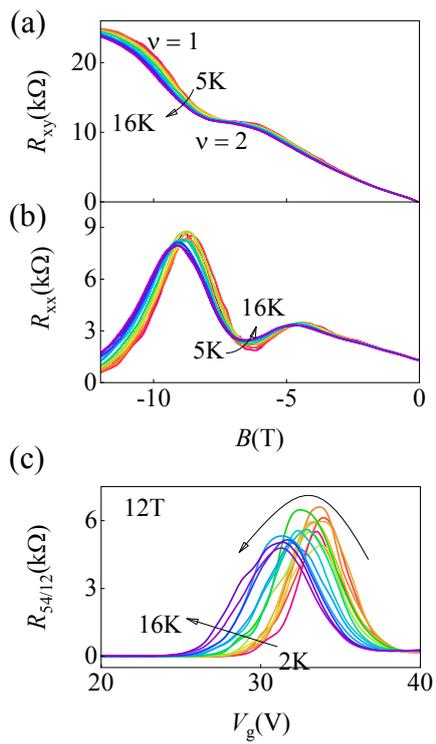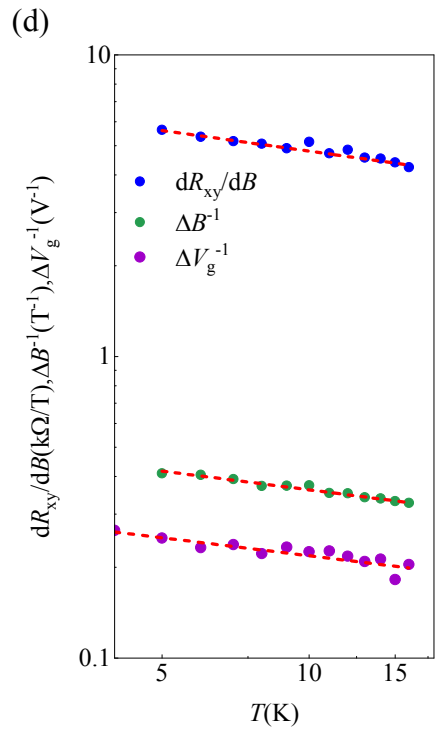

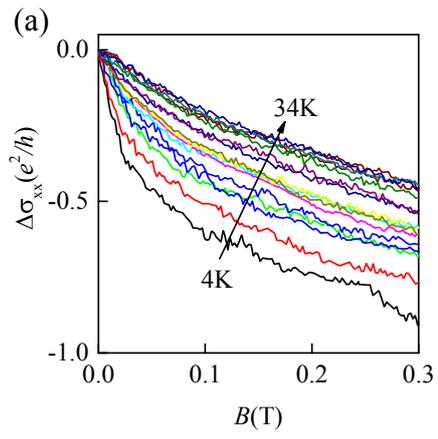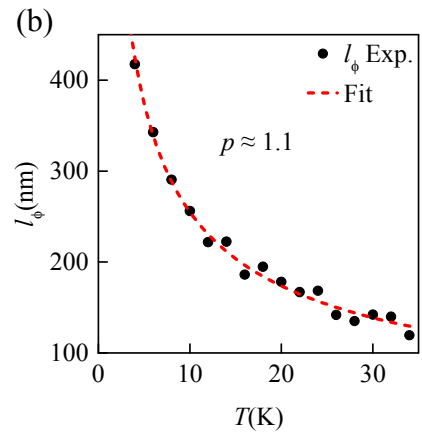